# General quasi-non-spreading linear three-dimensional wavepackets

Olga V. Borovkova, Yaroslav V. Kartashov, Valery E. Lobanov, Victor A. Vysloukh, and Lluis Torner

*ICFO-Institut de Ciencies Fotoniques, and Universitat Politecnica de Catalunya, Mediterranean Technology Park, 08860 Castelldefels (Barcelona), Spain*
*Corresponding author: Yaroslav.Kartashov@icfo.es*

We introduce a general approach for generation of sets of three-dimensional quasi-non-spreading wavepackets propagating in linear media, also referred to as linear light bullets. The spectrum of rigorously non-spreading wavepackets in media with anomalous group velocity dispersion is localized on the surface of a sphere, thus drastically restricting the possible wavepacket shapes. However, broadening slightly the spectrum affords the generation of a large variety of quasi-non-spreading distributions featuring complex topologies and shapes in space and time that are of interest in different areas, such as biophysics or nanosurgery. Here we discuss the method and show several illustrative examples of its potential. © 2011 Optical Society of America

Wavepackets that do not spread in time and/or space upon propagation in linear optical media are of paramount importance in diverse applications, such as medical imaging, microscopy, tomography, lithography, data storage, interconnects, optical tweezing and trapping, and optical lattices, to name a few [1,2]. The shapes of non-spreading field distributions depend on their dimensionality. For example, in two-dimensional geometries non-diffracting patterns can be constructed in the coordinate systems where the Helmholtz equation is separable, yielding solutions invariant upon propagation, such as plane waves in Cartesian coordinates, Bessel beams in circular cylindrical coordinates [3], Mathieu beams in elliptic cylindrical coordinates [4], and parabolic beams in parabolic cylindrical coordinates [5]. Direct engineering of the spatial spectrum allows construction of more sophisticated patterns (see [6] and references therein).

Especially interesting for practical applications is the generation of fully three-dimensional non-spreading wavepackets [7-12]. Experimental demonstrations of such wavepackets for normal and anomalous dispersion have been reported [13-15]. Recently, the propagation of accelerating Airy-Bessel wavepackets was observed [16] and beams featuring Airy shapes in three dimensions were generated [17] in a landmark advance. Such wavepackets were termed linear light bullets, by analogy with their more elusive self-sustained counterparts that exist in suitable nonlinear media [18,19]. Non-spreading three-dimensional wavepackets may be important in applications where propagation of a focused beam of short pulsed light over a significant depth of focus is crucial (e.g. in nanolithography and nanosurgery). Some applications may require sculptured complex spatial patterns beyond the known polychromatic combinations of Bessel beams. In this Letter we introduce an approach to generate such three-dimensional quasi-non-spreading wavepackets, with a variety of desired shapes. The method is based on the proper engineering of the spatiotemporal spectrum. The method allows generating shapes that have no analogs among the solutions known to date. They propagate undistorted over considerable distances, far more than those required in most of practical applications.

We describe the propagation of a three-dimensional linear wavepacket along the $\xi$ axis of a uniform medium with anomalous group velocity dispersion by the linear Schrödinger equation for the dimensionless light field amplitude $q$:

$$i\frac{\partial q}{\partial \xi} = -\frac{1}{2}\left(\frac{\partial^2 q}{\partial \eta^2} + \frac{\partial^2 q}{\partial \zeta^2} + \frac{\partial^2 q}{\partial \tau^2}\right), \quad (1)$$

where $\eta, \zeta$ are the normalized transverse coordinates, $\tau$ is the normalized retarded time, $\xi$ is the propagation distance. The profile of any non-spreading wavepacket that satisfies Eq. (1) and that propagates parallel to the $\xi$ axis can be presented in terms of the Whittaker integral [3-6], generalized to the case of three dimensions:

$$q_s = \exp(-ik_t^2 \xi / 2) \int_0^\pi d\theta \int_0^{2\pi} G(\theta,\varphi) \times \\ \exp[ik_t(\eta \cos\varphi \sin\theta + \zeta \sin\varphi \sin\theta + \tau \cos\theta)]d\varphi, \quad (1)$$

where $k_t = (k_\eta^2 + k_\zeta^2 + k_\tau^2)^{1/2}$ is the transverse wavenumber, $\varphi, \theta$ are the azimuthal and polar angles in the frequency space, respectively, and $G(\theta,\varphi)$ is the angular spectrum. The Fourier transform $q_k$ of the field $q_s$ shows that the angular spectrum $G(\theta,\varphi)$ of a rigorously non-spreading wavepacket is localized on the surface of a sphere of radius $k_t$. Physically, this means that only plane waves having equal transverse wavenumbers $k_t$ (hence, equal phase shifts accumulated upon propagation) may be involved into the formation of a non-spreading wavepacket. The wavenumber $k_t$ determines the characteristic transverse scale of the wavepacket, such as the radii of separate bright spots.

Integral (2) is the key ingredient for the construction of quasi-non-spreading wavepackets. Broadening the angular spectrum, i.e. using of spectral components with different transverse wavenumbers $k_t$ belonging to the finite spherical layer of width $\delta k_t$, dramatically enriches the shapes that can be generated. Naturally, such broadening causes wavepackets to spread slowly. However, the spreading remains slow as long as $\delta k_t \ll k_t$. In order to construct quasi-non-spreading wavepackets we used an iterative procedure which starts by selecting the desired field distribution $q_s(\eta,\zeta,\tau)$ at $\xi=0$ and calculating its Fourier spectrum $q_k(k_\eta,k_\zeta,k_\tau)$. Then, the spectrum is modified by setting to zero the amplitudes of spectral components $q_k$ with $(k_\eta^2+k_\zeta^2+k_\tau^2)^{1/2}$ values outside a finite spherical layer $[k_t-\delta k_t/2, k_t+\delta k_t/2]$. After the calculation of the inverse Fourier transform, one keeps new phase distribution $\arg[q_s(\eta,\zeta,\tau)]$, but replaces field modulus with the initial desired field modulus distribution. The iterative procedure is repeated until the convergence in phase is achieved [6].

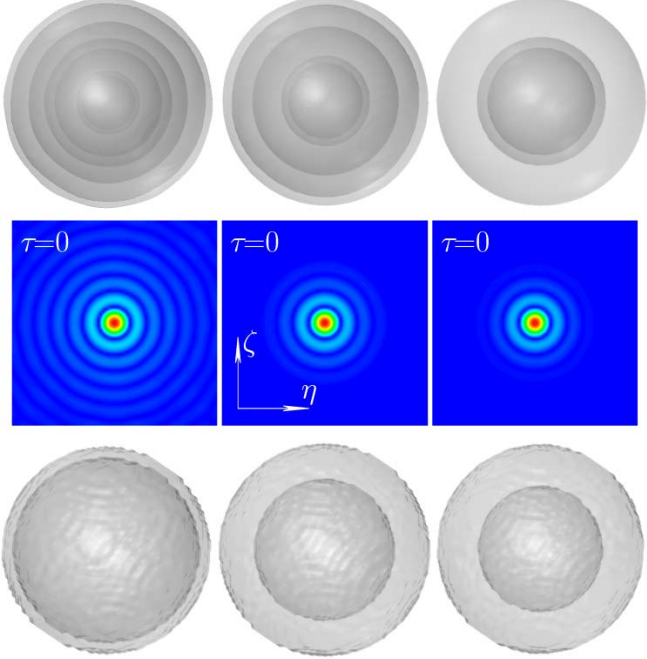

Figure 1. Isosurface plots at the level $0.08\max|q_s|$ showing 3D field modulus distributions (top row), field modulus distributions in the plane $\tau=0$ (middle row), and corresponding spectra at the level $0.4\max|q_k|$ (bottom row) of 3D wavepackets generated at $\delta=0.1$ (left column), $0.4$ (middle column), and $0.5$ (right column). All isosurface plots are made partially transparent to show internal structure of the wavepacket and its spectrum.

The method is illustrated in Fig. 1, where we show wavepackets constructed with different widths of the angular spectrum using as a trial function a localized Gaussian distribution (further we use $k_t=1$). For large relative widths of the spectrum $\delta=\delta k_t/k_t$ the resulting shape reproduces almost any trial function because multiple spectral components are involved into its formation (see Fig. 1, right, where at $\delta\sim 0.5$ the convergence towards a Gaussian distribution is apparent). However, such a wavepacket quickly spreads due to accumulated dephasing of spectral components. Decreasing the spectrum width results in the appearance of multiple rings around the bright central core (Fig. 1, left). These rings ensure non-spreading propagation by providing the transverse energy flow from the wavepacket periphery to its center. In the limit $\delta\to 0$ the method produces exact non-spreading wavepackets $q_s=[\sin(k_t\rho)/\rho]\exp(-ik_t^2\xi/2)$, where $\rho=(\eta^2+\zeta^2+\tau^2)^{1/2}$. The impact of the width of the angular spectrum on the rate of spreading is illustrated in Fig. 2(a). We defined $L$ as a propagation distance at which the peak amplitude decreases by 10%. When $\delta\to 0$, $L$ diverges, and when $\delta\sim 1$ (i.e., the width of the spectrum $\delta k_t$ is comparable to the transverse wavenumber $k_t$) the distance $L$ approaches the spreading length for the trial beam. Note that even for $\delta\sim 0.2$ the spreading length for the iterated wavepackets is sufficiently large that allows considering them non-spreading for most practical purposes. The broader the spectrum the richer the possible quasi-non-spreading patterns. Since exact non-spreading wavepackets carry an infinite energy, the fraction of power carried by the central bright spot decreases as $\delta\to 0$ [Fig. 2(b)].

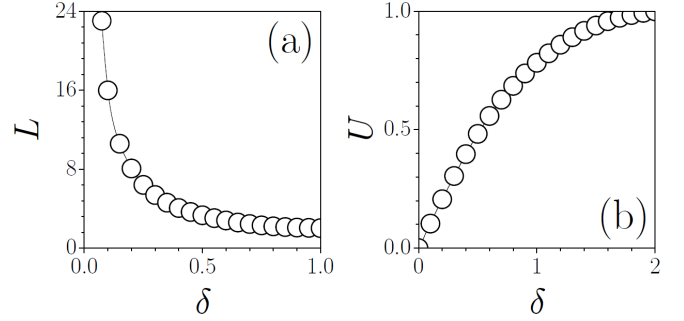

Figure 2. (a) The propagation distance at which peak amplitude of the wavepacket from Fig. 1 decreases by 10% versus angular spectrum width. (b) The fraction of energy concentrated in the central bright spot versus angular spectrum width.

The method allows formation of quasi-non-spreading wavepackets featuring radially symmetric or azimuthally modulated spatial distributions at any $\tau$ value and pronounced localization in time (Fig. 3). For example, Fig. 3 (left) shows a sequence of two wavepackets separated by a time interval and having Bessel-like spatial distributions. The finite width of angular spectrum allows to achieve a situation when temporal localization of each wavepacket is more pronounced than its spatial localization, as one can see from the cross-section at $\eta=0$. The spectrum of such a wavepacket resembles a finite-width ring wrapped around a sphere. One can generate three-dimensional wavepackets with spatial shapes corresponding to any known two-dimensional non-diffracting beam, such as Bessel, Mathieu, or parabolic beam. The temporal distributions of such wavepackets consist of the central bright lobe surrounded by multiple decaying oscillations. Arbitrary sequences of temporarily separated or strongly overlapping Bessel, Mathieu, or parabolic wavepackets can be generated too.

A specific necklace-like wavepacket that at each point of time resembles a higher-order azimuthally modulated Bessel beam is shown in Fig. 3 (center). Such a wavepacket is characterized by more localized field modulus distribution in $\tau=0$ plane than the azimuthally modulated Bessel beam. Interestingly, the spectrum of this beam is also azimuthally modulated. The number of nodes in spectrum corresponds to the number of azimuthal nodes in spatial distributions. One can generate wavepackets localized in time with sophisticated spatial profiles, such as necklace-like wavepackets shown in Fig. 3 (right) that exhibit different number of azimuthal nodes at $\eta<0$ and $\eta>0$ at any time moment $\tau$.

Distributions that have no analogs among three-dimensional patterns known to date are illustrated in Fig. 4. Such wavepackets feature complex shapes and may change topology between different temporal and spatial cross-sections. Examples include hollow three-dimensional shapes (Fig. 4, left) whose field modulus distribution in any crosssection $\eta=0$, $\zeta=0$, or $\tau=0$ resembles a ring surrounding the region with very small field intensity. The spectrum of such beam is composed of several rings with gradually decreasing widths that wrap around the sphere. One can generate a sequence of wavepackets which are well localized in space and whose centers may follow complex shapes in time. An example that exhibits spiraling in time is illustrated in Fig. 4 (center). Finally, a specific spherically periodic wavepacket is shown in Fig. 4 (right). Such wavepacket shows remarkable field periodicity at the periphery, while the field amplitude does not decrease at $\rho\to\infty$ as it happens in truly nondiffracting spherically symmetric beam with $q_s\sim\sin(k_t\rho)/\rho$.

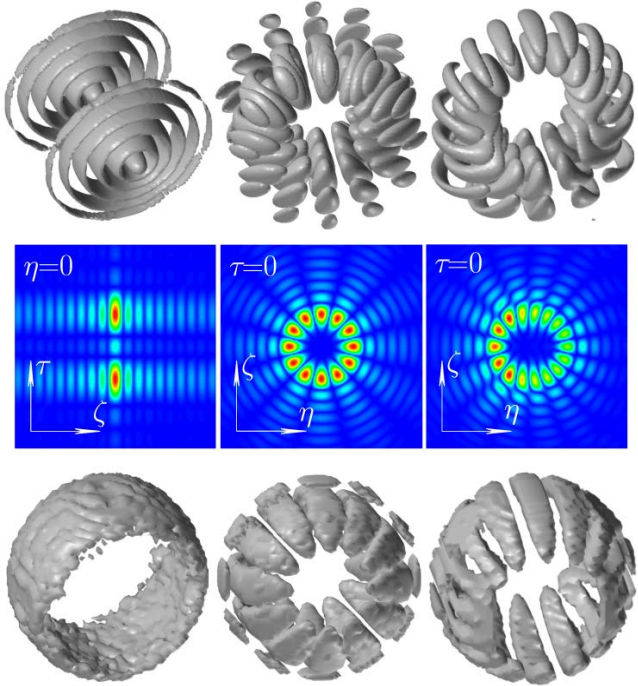

Figure 3. (Left) Field modulus distribution at the level $0.19\max|q_s|$ (top row), field modulus distribution at $\eta=0$ (middle row), and beam spectrum at the level $0.02\max|q_k|$ (bottom row) for wavepacket composed of two Bessel-like pulsed beams. (Center) Field modulus distribution at the level $0.19\max|q_s|$ (top row), field modulus distribution at $\tau=0$ (middle row), and spectrum at the level $0.14\max|q_k|$ for necklace-like wavepacket. (Right) Field modulus distribution at the level $0.17\max|q_s|$ (top row), field modulus distribution at $\tau=0$ (middle row), and spectrum at $0.36\max|q_k|$ for necklace wavepacket with complex spatial distribution.

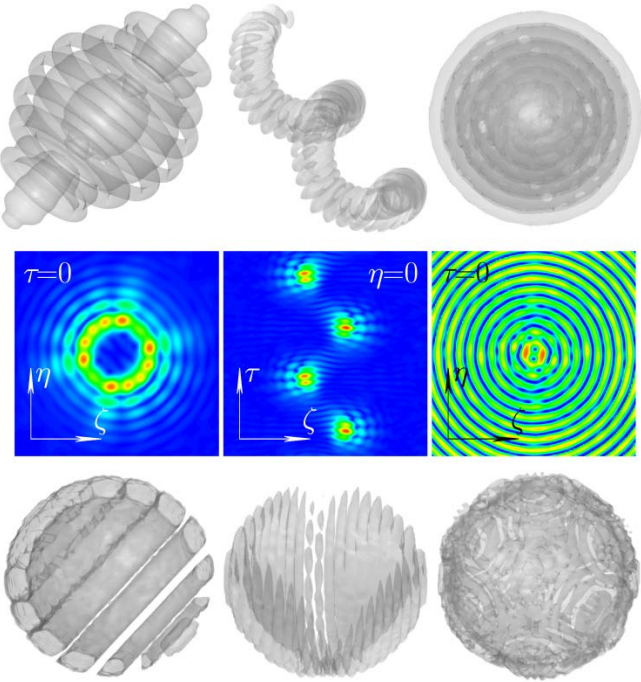

Figure 4. (Left) Field modulus distribution at the level $0.13\max|q_s|$ (top row), field modulus distribution at $\tau=0$ (middle row), and beam spectrum at the level $0.02\max|q_k|$ (bottom row) for "hollow" wavepacket. (Center) Field modulus distribution at the level $0.20\max|q_s|$ (top row), field modulus distribution at $\eta=0$ (middle row), and spectrum at the level $0.05\max|q_k|$ for spiraling wavepacket. (Right) Field modulus distribution at the level $0.20\max|q_s|$ (top row), field modulus distribution at $\tau=0$ (middle row), and spectrum at the level $0.16\max|q_k|$ for spherically periodic wavepacket (only several central rings are shown).

We conclude by stressing that the method used here allows the generation of a large variety of quasi-non-spreading patterns, beyond the exact solutions obtained from the Helmholtz equation in separable coordinates. Such quasi-non-spreading patterns may be tailored to meet the requirements of each particular application. Potential applications where such a sculptured pulsed-beams may be crucial are anticipated in areas where tight focus, large depth of focus and high peak power are required.